\documentclass{article}

\usepackage{amsfonts,amssymb,amsmath} 
\usepackage{arxiv}
\usepackage{bmpsize}
\usepackage{booktabs}                 
\usepackage{braket}
\usepackage{caption}
\usepackage{circuitikz}
\usepackage{colortbl}
\usepackage{doi}
\usepackage{graphicx}
\usepackage{epstopdf}
\usepackage{lipsum}		              
\usepackage{mathtools}
\usepackage{microtype}                
\usepackage{natbib}
\usepackage{nicefrac}                 
\usepackage{natbib}
\usepackage{tabulary,xcolor}
\usetikzlibrary{angles,quotes}
\usepackage{subfigure}
\usepackage{pdflscape}
\usepackage{pifont}
\usepackage{url,multirow,morefloats,floatflt,cancel,tfrupee}
\usepackage[nointegrals]{wasysym}
\usepackage{xcolor}

\title{Quantum Computing: Fundamentals, Trends and Perspectives for Chemical and Biochemical Engineers}

\author{{Amirhossein Nourbakhsh}\\
	Geomatics Engineering\\
	Lassonde School of Engineering\\
	York University\\
	Toronto, ON M3J 1P3\\
	Canada
	\And
	{Mark Nicholas Jones}\thanks{mark@mqs.dk}\\
	Molecular Quantum Solutions\\
	Maskinvej 5\\
	2860 Søborg\\
	Denmark\\
	\And
	{Kaur Kristjuhan}\\
	Molecular Quantum Solutions\\
	Maskinvej 5\\
	2860 Søborg \\
	Denmark
	\And
	{Deborah Carberry}\\
	Department of Chemical\\and Biochemical Engineering\\
	Technical University of Denmark\\
	Søltofts Plads\\
	2800 Kongens Lyngby\\
	Denmark
	\And
	{Jay Karon}\\
	Intelligent Data Analytics\\
	Toronto, ON M3J 1P3\\
	Canada
	\And
	{Christian Beenfeldt}\\
	Knowledge Hub Zealand\\
	Biotech City Kalundborg\\
	4400 Kalundborg\\
	Denmark
	\And
	{Kyarash Shahriari}\\
	Intelligent Data Analytics\\
	Toronto ON M3J 1P3\\
	Canada
	\And
	{Martin P Andersson}\\
	Department of Chemical\\and Biochemical Engineering\\
	Technical University of Denmark\\
	Søltofts Plads\\
	2800 Kongens Lyngby\\
	Denmark
	\And
	{Mojgan A. Jadidi}\thanks{mjadidi@yorku.ca}\\
	Geomatics Engineering\\
	Lassonde School of Engineering\\
	York University\\
	Toronto, ON M3J 1P3\\
	Canada
	\And
	{Seyed Soheil Mansouri}\thanks{seso@kt.dtu.dk}\\
	Department of Chemical\\and Biochemical Engineering\\
	Technical University of Denmark\\
	Søltofts Plads\\
	2800 Kongens Lyngby\\
	Denmark
}

\renewcommand{\headeright}{Preprint}
\renewcommand{\undertitle}{Preprint}
\renewcommand{\shorttitle}{\textit{arXiv}}

\begin{document}
\maketitle

\pagebreak

\begin{abstract}
We use the benefits and components of classical computers every day. However, there are many types of problems which, as they grow in size, their computational complexity grows larger than classical computers will ever be able to solve. Quantum computing (QC) is a computation model that uses quantum physical properties to solve such problems. QC is at the early stage of large-scale adoption in various industry domains to take advantage of the algorithmic speed-ups it has to offer. It can be applied in a variety of areas, such as computer science, mathematics, chemical and biochemical engineering, and the financial industry. The main goal of this paper is to give an overview to chemical and biochemical researchers and engineers who may not be familiar with quantum computation. Thus, the paper begins by explaining the fundamental concepts of QC.
The second contribution this publication tries to tackle is the fact that the chemical engineering literature still lacks a comprehensive review of the recent advances of QC. Therefore, this article reviews and summarizes the state of the art to gain insight into how quantum computation can benefit and optimize chemical engineering issues.
\\
A bibliography analysis covers the comprehensive literature in QC and analyzes quantum computing research in chemical engineering on various publication topics, using Clarivate analytics covering the years 1990 to 2020.
After the bibliographic analysis, relevant applications of QC in chemical and biochemical engineering are highlighted and a conclusion offers an outlook of future directions within the field.
\end{abstract}


\section{Introduction}
The term quantum computer was first coined by Paul Benioff in 1979 when he constructed a microscopic quantum mechanical model to perform computation. \citep{benioff1980computer} In the early 70's, Stephen Wiesner proposed his idea of “Conjugate Coding,” which is a cryptographic tool, and submitted an article to the IEEE Transactions on Information Theory. His paper was rejected because it seemed incomprehensible to the referees. \citep{brassard2005brief} In October 1979, Benioff and Wiesner found a way to use Wiesner’s coding scheme with the concept of public-key cryptography. \citep{diffie1976new} Finally, the result of their collaboration led to the generation of new topics in the field, such as teleporting an unknown quantum state by using dual classical and Einstein-Podolsky-Rosen channels \citep{bennett1993teleporting}, entanglement distillation \citep{bennett1996purification}, strengths and weaknesses of quantum computing \citep{bennett1997strengths}, and quantum cryptography \citep{bennett1992quantum}.

Richard Feynman's description of how to accurately calculate quantum physical system on a quantum computer architecture can be seen as the first description of a quantum computer architecture.
He submitted the description on the 7th of May 1981 to International Journal of Theoretical Physics. The article was then published in volume 21, June, 1982.

David Deutsch described in 1985 his idea of an universal quantum computer. He mentioned that a computing machine with quantum theory characteristics can be built and that it can have many remarkable properties which cannot be found in Turing machines. \citep{deutsch1985quantum}

\begin{figure}[htbp!]
\centering
  \includegraphics[width=1.0\textwidth]{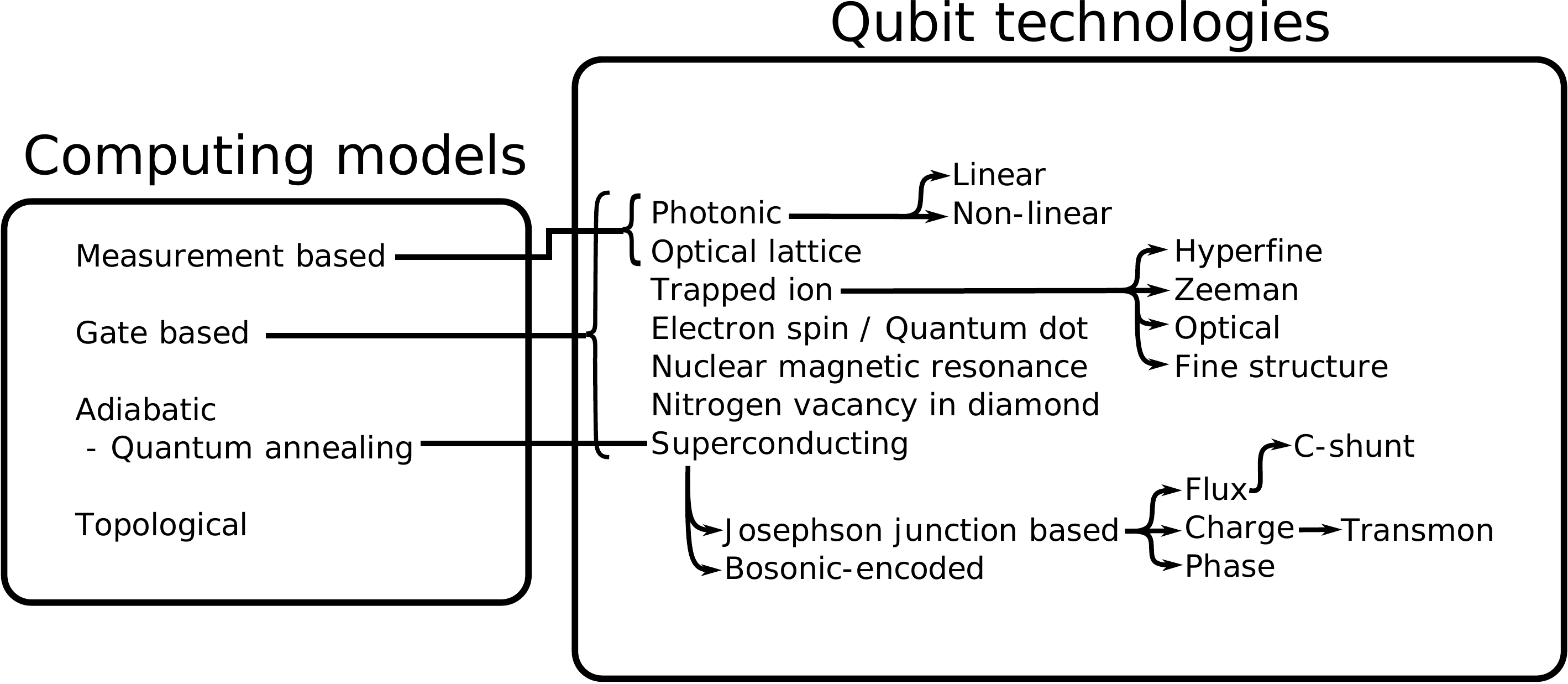}\\
  \caption{A multi-layered view of quantum computing containing Qubit technologies and quantum computation models}\label{figView1}
\end{figure}     

Quantum computing has been studied for decades and ranges from information theory to details of hardware technologies, computational models implemented on the specific hardware and problem formulations which can be solved on a certain hardware type.
It has applications in nearly every field that contains or utilizes computations with high complexity. Quantum computers have the potential to impact many aspects of current domains of science, including computer science, mathematics, and chemical engineering. Generally, to compare classical computing and quantum computing, computational problems can be divided into three categories. The first category contains problems that cannot be solved by classical computers in any feasible amount of time, but which are possible to solve with quantum computers. This is one of the main goals of quantum computing, which is termed “quantum supremacy" or "quantum advantage".
John Preskill published a paper in November 2012 about quantum supremacy and the consequences it will have to several critical applications in society such as: cryptography and optimization. \citep{preskill2012quantum}
\\
In October 2019, Google claimed to have achieved quantum supremacy using a programmable superconducting processor. \citep{arute2019quantum} Google scientists used a quantum processor called “Sycamore” to sample the output of a pseudo-random quantum circuit. They used 53 qubits representing a state space of \(2^{53}\) dimensions. The process took about 200 seconds to solve a complex computation, while that process would have taken about 10,000 years with a classical computer. \citep{arute2019quantum} However, IBM stated that the computation of the Google experiment could be performed on a classical computer in 2.5 days rather than 10,000 years. \citep{edwin2019} The second category of computational problems is related to the issues that can be solved with both classical computers and quantum computers, though with less computational complexity when using a quantum computer. The third category is comprised of problems that cannot be solved more efficiently with quantum computers. Thus, a careful analysis of the mathematical complexity of a mathematical problem formulation must be made to assess which computational system/architecture should be applied.
\\
Figure \ref{figView1} represents which Qubit technologies enable which kind of quantum computing model. Each provider has its model of QC technology. The details of the different QC technologies are described in Section 2 and 3.  

\section{Fundamentals of Quantum Computing}
There are four computation models in quantum computers:
\begin{itemize}
\item[] 1. gate model (circuit)
\item[] 2. adiabatic
\item[] 3. measurement based
\item[] 4. topological.
\end{itemize}
In the following sections, these models are described.

\subsection{Gate-based Quantum Computing}
\begin{figure}[htbp!]
\centering
  \begin{tikzpicture}

    \def\r{2}

    \draw [-latex, line width=0.4mm] (0,0) node[circle,fill,inner sep=1] (orig) {} -- (\r/3,\r/2) node[label=$\ket{\psi}$] (a) {};
    \draw[dashed] (orig) -- (\r/3,-\r/5) node (phi) {} -- (a);

    \draw (orig) circle (\r);
    \draw[dashed] (orig) ellipse (\r[0] and \r/3);

    \draw[dashed] (orig) -- ++(-\r/5,-\r/3) node[below] (x1) {};
    \draw[dashed] (orig) -- ++(\r,0) node[right] (x2) {};
    \draw[dashed] (orig) -- ++(0,\r) node[above] (x3) {$\ket{0}$};
    \draw[dashed] (orig) -- ++(0,-\r) node[below] (south) {$\ket{1}$};

    \pic [draw=gray,text=gray,-latex,"$\phi$"] {angle = x1--orig--phi};
    \pic [draw=gray,text=gray,latex-,"$\theta$"] {angle = a--orig--x3};

    \end{tikzpicture}
  \caption{The state $\ket{\psi}$ of a qubit on the surface of the Bloch sphere by using two polar coordinates $\theta$ and $\phi$. The poles correspond to the states $\ket{0}$ and $\ket{1}$ by definition.}\label{figBloch}
\end{figure}
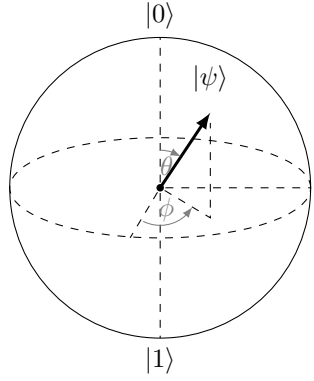

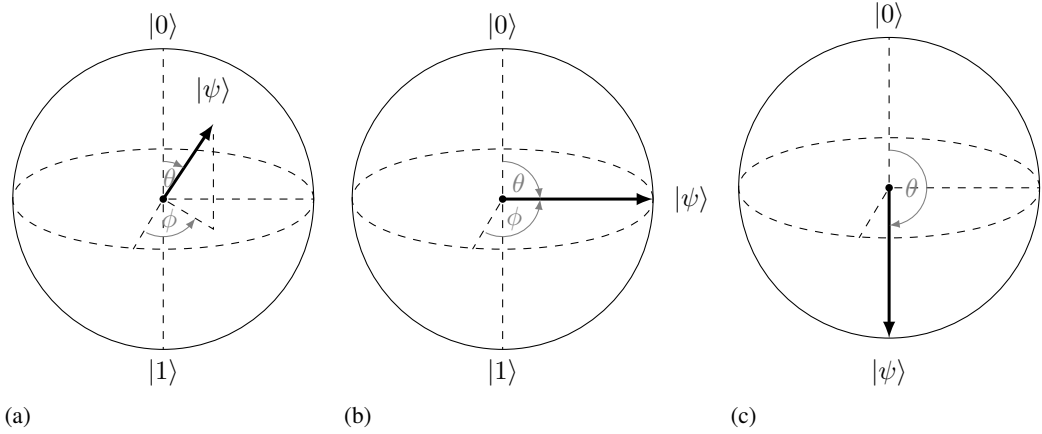
\begin{figure}[htbp!]
\centering
\begin{subfigure}[]{}
    \begin{tikzpicture}

    \def\r{2}

    \draw [-latex, line width=0.4mm] (0,0) node[circle,fill,inner sep=1] (orig) {} -- (\r/3,\r/2) node[label=$\ket{\psi}$] (a) {};
    \draw[dashed] (orig) -- (\r/3,-\r/5) node (phi) {} -- (a);

    \draw (orig) circle (\r);
    \draw[dashed] (orig) ellipse (\r[0] and \r/3);

    \draw[dashed] (orig) -- ++(-\r/5,-\r/3) node[below] (x1) {};
    \draw[dashed] (orig) -- ++(\r,0) node[right] (x2) {};
    \draw[dashed] (orig) -- ++(0,\r) node[above] (x3) {$\ket{0}$};
    \draw[dashed] (orig) -- ++(0,-\r) node[below] (south) {$\ket{1}$};

    \pic [draw=gray,text=gray,-latex,"$\phi$"] {angle = x1--orig--phi};
    \pic [draw=gray,text=gray,latex-,"$\theta$"] {angle = a--orig--x3};

    \end{tikzpicture}
\end{subfigure}
\begin{subfigure}[]{}
    \begin{tikzpicture}

    \def\r{2}

    \draw [-latex, line width=0.4mm] (0,0) node[circle,fill,inner sep=1] (orig) {} -- (\r,0) node[label=right:$\ket{\psi}$] (a) {};
    \draw (orig) circle (\r);
    \draw[dashed] (orig) ellipse (\r[0] and \r/3);

    \draw[dashed] (orig) -- ++(-\r/5,-\r/3) node[below] (x1) {};
    \draw[dashed] (orig) -- ++(0,\r) node[above] (x3) {$\ket{0}$};
    \draw[dashed] (orig) -- ++(0,-\r) node[below] (south) {$\ket{1}$};

    \pic [draw=gray,text=gray,-latex,"$\phi$"] {angle = x1--orig--a};
    \pic [draw=gray,text=gray,latex-,"$\theta$"] {angle = a--orig--x3};

    \end{tikzpicture}
\end{subfigure}
\begin{subfigure}[]{}
    \begin{tikzpicture}

    \def\r{2}

    \draw [-latex, line width=0.4mm] (0,0) node[circle,fill,inner sep=1] (orig) {} -- (0,-\r) node[label=below:$\ket{\psi}$] (a) {};

    \draw (orig) circle (\r);
    \draw[dashed] (orig) ellipse (\r[0] and \r/3);

    \draw[dashed] (orig) -- ++(-\r/5,-\r/3) node[below] (x1) {};
    \draw[dashed] (orig) -- ++(\r,0) node[right] (x2) {};
    \draw[dashed] (orig) -- ++(0,\r) node[above] (x3) {$\ket{0}$};
    \pic [draw=gray,text=gray,latex-,"$\theta$"] {angle = a--orig--x3};

    \end{tikzpicture}
\end{subfigure}
  \caption{Visualization of 3 different qubit states. Figure (a) is described with $\theta=30^o$ and $\phi=60^o$. In Figure (b), the state $\ket{\psi}$ is described by $\phi=\theta=90^o$ and is in an equal superposition of $\ket{0}$ and $\ket{1}$. In Figure (c), the state $\ket{\psi}=\ket{1}$ has no defined value of $\phi$, because the first term in equation (\ref{eq9}) is zero for $\theta=180^o$, making $\phi$ part of the global phase variable $\phi_0$, which was neglegted after equation (\ref{eq8})}
  \label{figBloch2}
\end{figure}

Superposition and entanglement are one of the features of quantum physics that quantum computing makes use of and are being introduced in the following. In classical computers a bit can only be in one of two states, usually represented as 1 and 0. However, a qubit, the basic building block of QC, can be in a superposition state of 1 and 0. Another difference between a bit and a qubit is that measuring a bit has no effect on its state. However, the measurement of a qubit changes its state, except if already being in the computational basis state of 0 and 1. Moreover, a classic system, given the same inputs, will always yield the same results. However, QCs with the same system and inputs will give different results with a certain probability. \citep{west2000quantum}
\\
There exist different theoretical approaches to depict a state of a qubit, many times related to the technological implementation of a qubit. One of the well-known ways to depict a qubit state is by utilizing the Bloch sphere diagram. As Figure \ref{figBloch} shows, every qubit state can be described in polar coordinates by two angle parameters, \(\theta\) and \(\phi\), which give us a point on the sphere; the zero state \(\ket{0}\) is located on the north pole of the Bloch sphere and the one state \(\ket{1}\) is located on the south pole.
\\
\\
This allows to visualize 2-dimensional states in the 3-dimensional space. In other words, instead of considering a qubit as a 2-dimensional complex vector space on the unit sphere, we can equivalently consider the state of the qubit as sitting on the surface of a 3-dimensional unit sphere by introducing the two parameters \(\alpha\) and \(\beta\):

\begin{equation} \label{eq6}
\ket{\Psi} = \alpha\ket{0}+\beta\ket{1} 
\end{equation}
\\
where \(\ket{\Psi}\) represents the state of the qubit and \(\alpha\) and \(\beta\) are complex numbers and \(|\alpha|^2 + |\beta|^2 = 1\). Based on the definition of complex vectors in polar coordinate systems the qubit state can be expressed as:

\begin{equation} \label{eq7}
\begin{array}{ll}
\ket{\Psi} = r_0e^{i\phi_0}\ket{0} + r_1e^{i\phi_1}\ket{1} , & r_0^2+r_1^2=1

\end{array}
\end{equation}
\\
which leads to: 
\\
\begin{equation} \label{eq8}
\ket{\Psi} = e^{i\phi_0}[r_0\ket{0} + r_1e^{i(\phi_1 - \phi_0)} \ket{1}]
\end{equation}
\\
The term \(e^{i\phi_0}\) does not change the result after each measurement of the system and can therefore be neglected.
Moreover,  \(r_0\) and \(r_1\) can be replaced by \(\cos{\frac{\theta}{2}}\) and \(\sin{\frac{\theta}{2}}\). Thus, the following equation describes the general state of the qubit on the Bloch sphere by using two parameters in 3-dimensions.

\begin{equation} \label{eq9}
\ket{\Psi} = \cos{\frac{\theta}{2} \ket{0}} + e^{i\phi} \sin{\frac{\theta}{2}} \ket{1}
\end{equation}
\\
The variables \(\theta\) and \(\phi\) parametrize the entire Bloch sphere, and every possible quantum state of a single qubit can be expressed by formula \eqref{eq9}. A selection of quantum states is illustrate in Figure \ref{figBloch2}.

To change the state of a qubit, quantum gates are applied. There are many kinds of quantum gates as well as many different physical implementations of them. Some of them are basic gates and can transform one or two qubits at the moment. Others are constructed from specific arrangements of several basic quantum gates, called compound gates.

Any movement on the Bloch sphere is spanned by the 3-dimensional rotation group $SO(3)$, which is isomorphic to the special unitary group $SU(2)$. All elements of this group can be represented as 2-dimensional unitary matrices with unit determinant. In this representation, the qubit basis states are represented as vectors:

\begin{equation} \label{eq91}
\ket{0} = \left[\begin{array}{c}1\\0\end{array}\right]\quad\ket{1} = \left[\begin{array}{c}0\\1\end{array}\right]
\end{equation}

A selection of commonly used single qubit gates are represented in Table \ref{table1}.

\begin{table}[htbp!]
\caption{Single qubit gates with their matrices, their effect on the $\ket{0}$ state and a general description.}
\label{table1}
\resizebox{\linewidth}{!}{%
\begin{tabular}{c|c|c|c|c}
Gate & Matrix  & Input & Output & Description \\
\hline
Pauli-X & $
\begin{bmatrix}
0 & 1\\
1 & 0
\end{bmatrix} $ & $ \ket{0} $ & $ \ket{1} $ & Performs a rotation of $\pi$ around the $X$ axis. Also called the NOT gate.  \\
\hline
Pauli-Y & $ \begin{bmatrix}
0 & -i\\
i & 0
\end{bmatrix} $ & $ \ket{0} $ & $ i\ket{1} $ & Performs a rotation of $\pi$ around the $Y$ axis. \\
\hline
Pauli-Z & $ \begin{bmatrix}
1 & 0\\
0 & -1
\end{bmatrix} $ & $ \ket{0} $ & $ \ket{0} $ & Performs a rotation of $\pi$ around the $Z$ axis. \\
\hline
Hadamard & $ \frac{1}{\sqrt{2}}\begin{bmatrix}
1 & 1\\
1 & -1
\end{bmatrix} $ & $ \ket{0} $ & $ \frac{1}{\sqrt{2}}\left(\ket{0}+\ket{1}\right) $ & Performs a rotation of $\pi$ around the $Z$ axis, followed by a rotation of $\pi/2$ around the $Y$ axis\\
\hline
$ R_\phi $ & $ \begin{bmatrix}
1 & 0\\
0 & e^{i\phi}
\end{bmatrix} $ & $ \ket{0} $ & $ \ket{0} $ & Movement along the latitudal direction, by an angle of $\phi$. Also called shift gate, where the $\phi$ is the phase shift.  \\
\hline
I-gate & $ \begin{bmatrix}
1 & 0\\
0 & 1
\end{bmatrix} $ & $ \ket{0} $ & $ \ket{0} $ & Does not change the state. Also called the identity or no-op gate.\\
\hline
S-gate & $ \begin{bmatrix}
1 & 0\\
0 & i
\end{bmatrix} $ & $ \ket{0} $ & $ \ket{0} $ & Performs a phase shift of $\pi/2$  \\
\hline
T-gate & $ \begin{bmatrix}
1 & 0\\
0 & e^{\frac{i\pi}{4}}
\end{bmatrix} $ & $ \ket{0} $ & $ \ket{0} $ & Performs a phase shift of $\pi/4$. \\
\end{tabular}}
\end{table}

A quantum circuit can consist of several gates to construct a sequence operations on the available multi-qubit system. To perform a quantum computation with multiple qubits, the qubits must interact with each other in some manner. In the gate-based model of QC, this is achieved by introducing gates that act on multiple qubits. When a single qubit gate was representable as a $2$-dimensional unitary matrix, then an $n$ qubit gate is representable as a $2^n$-dimensional unitary matrix. The simplest and most commonly occuring example is a two qubit gate called the CNOT gate, given by the matrix

\begin{equation} \label{eq101}
\text{CNOT} = \left[\begin{array}{cccc}1&0&0&0\\0&1&0&0\\0&0&0&1\\0&0&1&0\end{array}\right]
\end{equation}

It is useful for creating entanglement between two qubits, meaning that it can create a two qubit state which cannot be expressed as a direct product of two single qubit states. An example is the two qubit Bell state

\begin{equation} \label{eq11}
\ket{\Psi} = \frac{1}{\sqrt{2}} \ket{00} + \frac{1}{\sqrt{2}} \ket{11}
\end{equation}

which can be prepared from the $\ket{00}$ state by applying a Hadamard gate on the first qubit, followed by a CNOT gate controlled by the first qubit and targeted on the second qubit. Both of the qubits are in an equal superposition of $\ket{0}$ and $\ket{1}$, but they are correlated with eachother - no matter the measurement outcome, we know that the two qubits will be in the same state.

The physical implementation of quantum gates is specific to each qubit technology. However, there is a universal trade-off that has to be considered by all of them. Namely, to apply gates to qubits, they must be susceptible to interactions, or else we cannot address them. Unfortunately, this also produces susceptibility to noise. This is especially problematic for entangled states, where each additional degree of entanglement will increase the amount of interference from noise.

\subsection{Adiabatic Model}
In this section, the adiabatic quantum computation model is described. The Hamiltonian is an operator that can be used to determine the total energy of a system and consists of kinetic and potential energy. Let \(N\) be the number of qubits available on the quantum hardware system. \(N\) qubits are set up in the ground state of \(H_0\). \(H_0\) is gradually transformed into a new Hamiltonian (\(H_f\)). In other words, \(H_0\) is the initial Hamiltonian, and \(H_f\) is the final Hamiltonian. A gradual transformation can be described with the following equation:

\begin{equation} \label{eq1}
H(t) = (1-t)H_0 + t H_f
\end{equation}

where \(t\) is a given time that gradually increased from 0 to 1 until \(H_f\) is reached. The ground state of \(H_f\) is \(\ket{\Psi_f}\), which is unknown. Finding \(\ket{\Psi_f}\) is the problem that adiabatic quantum computing is trying to solve. The quantum adiabatic theorem states that if one starts from the ground state of the inital hamiltonian and the transformation is done slowly enough, then the final state of the system is the ground state of the final hamiltonian. \citep{kato1950adiabatic}
One limitation is that the the minimal required time needed to perform the transformation scales as follows:

\begin{equation} \label{eq2}
T = \frac{1}{(Min (g(t)))^2}
\end{equation}

where \(g(t)\) is the difference between the two smallest eigenvalues of \(H(t)\) (the difference between the ground energy and the first excited energy). Adiabatic quantum computing has been shown to be capable of achieving a speedup over classical algorithms \cite{roland2002quantum}, where similar to Grover's algorithm \cite{grover1996fast}, finding any given element in an unstructured list could be done in \(O(\sqrt{N})\) steps (as opposed to the requisite \(O(N)\) for classical computers.

\subsubsection{Quantum Annealing}
Quantum annealing belongs to the category of adiabatic quantum computing and is a meta-heuristic algorithm, which is able to find the minimum of a discrete cost (energy) function. Not all optimization problems can be solved with this approach, but there exist physical implementations that are able to find the ground state of a specific spin-lattice model called the transverse Ising model \cite{kadowaki1998quantum}, where each of the spins on the lattice is represented by a qubit. Quantum annealing is performed by preparing the ground state of a certain configuration of the spin-lattice model, described by some parameters, and subsequently changing those parameters to adiabatically transform the model such that it has a Hamiltonian that corresponds to the energy function of interest. Classically, finding the minimum of a discrete energy function can be considered as a combinatorical optimization problem.

The Hamiltonian of a transverse Ising model is:
\begin{equation}
    \hat{H} = \sum h_i \hat{\sigma}_z^i - \sum_{ij} J_{ij} \hat{\sigma}_z^i \hat{\sigma}_z^j
\end{equation}
\\
where $i$ and $j$ enumerate the different qubits, $\hat{\sigma}_z$ are Pauli Z operators, \(h_i\) are called qubit biases and $J_{ij}$ are called coupling strengths.
By configuring a network of radio frequency superconducting quantum interference devices (RF-SQUIDs), the values of \(h\) and \(J\) can be configured and smoothly adjusted during operation. \cite{harris2010experimental}
To solve a problem with a discrete energy function $E(s)$, where $s\in\{-1,1\}$, a corresponding Hamiltonian $\hat{H}(\hat{\sigma})$ with the same functional form needs to be prepared on the quantum computer. This is done by defining a time-dependent hamiltonian, which adiabatically transforms an easily preparable Hamiltonian $\hat{H}_X(\hat{\sigma})$ into the problem Hamiltonian $\hat{H}(\hat{\sigma})$:
\begin{equation}
    \hat{H}(t, \hat{\sigma}) = A(t)\hat{H}_X(\sigma) + B(t)\hat{H}(\hat{\sigma})
\end{equation}
where $A(t)$ and $B(t)$ are functions that make up the so-called annealing schedule and $A(t)\gg B(t)$ at the start of annealing $t=t_\text{start}$ whereas $A(t)\ll B(t)$ at the end of annealing $t=t_\text{end}$. In the RF-SQUID based approach, the starting Hamiltonian has a form of
\begin{equation}
    \hat{H}_X(\sigma) = \sum_i \hat{\sigma}_x^i
\end{equation}
If the system is prepared in a ground state of $\hat{H}_X(\sigma)$ at $t=t_\text{start}$, then it will be in a ground state of $\hat{H}(\sigma)$ at $t=t_\text{end}$ if \(t\) changes sufficiently slowly in accordance with the adiabatic theorem provided before. Therefore measuring the states of the qubits will yield the ground state of the problem Hamiltonian, which represents the lowest-energy configuration of the original energy function, with high probability.

\subsubsection{Quadratic unconstrained binary optimization (QUBO)}
To make use of current quantum annealing hardware (e.g. from D-Wave or Atos), one has to formulate the problem at hand as either an Ising problem
\begin{equation}
    E_{Ising} = \sum h_i s^i - \sum_{ij} J_{ij} s^i s^j
\end{equation}
or a quadratic unconstrained binary optimization (QUBO) problem:
\begin{equation}
    E_{QUBO} = x^T Qx = \sum_{ij} Q_{ij}x_ix_j
\end{equation}
where x is a binary vector and Q is a matrix.
The energy function of a QUBO resembles the Ising formulation except that the binary variables are either 0 and 1 in comparison to the spin values of the Ising model being -1 and 1. The two models are easily convertable to each other, and the QUBO form is often simpler to handle mathematically. Although this format of problem input may seem restrictive, many important NP complexity class problems can be expressed in this manner \cite{lucas2014ising}.

\subsection{Measurement based model}
Measurement-based quantum computing (MBQC) is an approach to quantum computing, where the computation is implemented by performing a sequence of measurements on a quantum computer. This is possible because measurements change the state of the quantum computer, and can thus be used to emulate the action of quantum gates, enabling the processing of quantum information. In fact, universal quantum computation is possible within this framework, first shown by Raussendorf and Briegel in 2001, where they devised a so-called \textit{one-way quantum computer}, which could implement the universal gate set using single-qubit measurements. \citep{raussendorf2001one}

Although measurements can be used to create entanglement \cite{pfaff2013demonstration}, it is more often the case that measurements destroy entanglement instead. For example, when performing any single-qubit measurement, the state of the measured qubit must necessarily collapse to a pure state, thus removing any entanglement it had in relation to other qubits. Because of this feature, MBQC most often relies on preparing a highly entangled state, called a \textit{resource state}, on the quantum computer beforehand, where entanglement can be thought of as a resource, which is successively used up by the measurements needed to perform the computation. The specific design of a resource state has to be carefully considered, because it was discovered by Gross et al., that if the resource state is chosen randomly (from the Haar measure), then MBQC offers no speedup for computation. \cite{gross2009most} Unfortunately, finding a suitable resource state for a given computation is in general more difficult than the computation itself \cite{morimae2017finding}. Moreover, physical hardware and design constraints limit the possible resoruce states that can be efficiently prepared, as it is very challenging to develop the capability to prepare arbitrarily entangled states. Perhaps the one of the most simple examples of a resource state is a \textit{cluster state}, which consists of a regular lattice of entangled qubits, such as the square lattice used in the original one-way quantum computer. \cite{raussendorf2001one}

Another crucial aspect to keep in mind with the MBQC scheme, is that results of measurements are inherently random. As a consequence, the unitary transformations induced by those measurements are concatenated with a random Pauli operator called a byproduct operator. \cite{morimae2014measurement} To avoid an indeterministic calculation, this randomness can be mitigated by selecting the basis of each measurement based on the results of the measurements preceding it. This extra classical processing of information can be harmful insofar as it may cause a time-delay, which can lead to increased decoherence in the quantum computer.

MBQC is particularly appealing for quantum computing with photons, where creating highly entangled cluster states is possible and performing single-qubit measurements is immensely simpler than performing multi-qubit operations required for the gate-based model. \cite{larsen2019deterministic} Additionally, Broadbent, Fitzsimons and Kashefi discovered that MBQC can be used for blind quantum computation, where a client can outsource their computations to a server in a way where the server cannot decipher what computation is being performed \cite{broadbent2009universal}, which is especially useful for future applications that have a high demand for security and privacy. \cite{fitzsimons2017private}
For a more in-depth review of MBQC, see \cite{wei2021measurement,briegel2009measurement,browne2016one}.

\subsection{Topological model}
A common feature of all the previously described computational models is that information is encoded into qubits locally. Through decoherence, information is lost when qubits interact with their local environment. The time that this process takes is called the decoherence time and it limits the accuracy and length of calculations that can be performed on quantum computers.

\begin{figure}[htbp!]
\centering
  \includegraphics[width=0.45\textwidth]{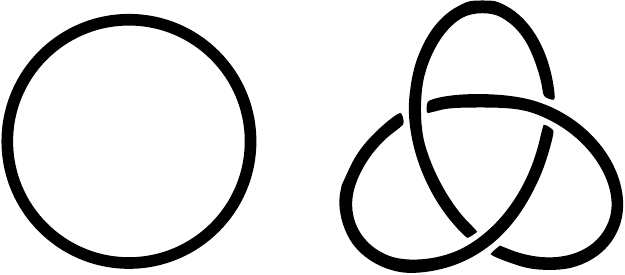}
  \caption{The trivial knot (left) and the left-handed trefoil knot(right). These knots are topologically inequivalent. If this difference is used as a basis for encoding information, then the information is protected from all rope deformations besides cutting.}\label{knots}
\end{figure}

The topological model of quantum computation proposes to circumvent decoherence by encoding information non-locally, into global characteristics known as topological invariants. A classical example of this, shown in Figure \ref{knots}, is encoding information by braiding ropes into knots that differ by their knot invariant. The two knots shown in the figure are topologically inequivalent, which means that one cannot be deformed into the the other without cutting the rope. This is a global property, since no particular part of either rope carries this information.

\begin{figure}[htbp!]
\centering
\begin{subfigure}[]{}
  \includegraphics[width=0.45\textwidth]{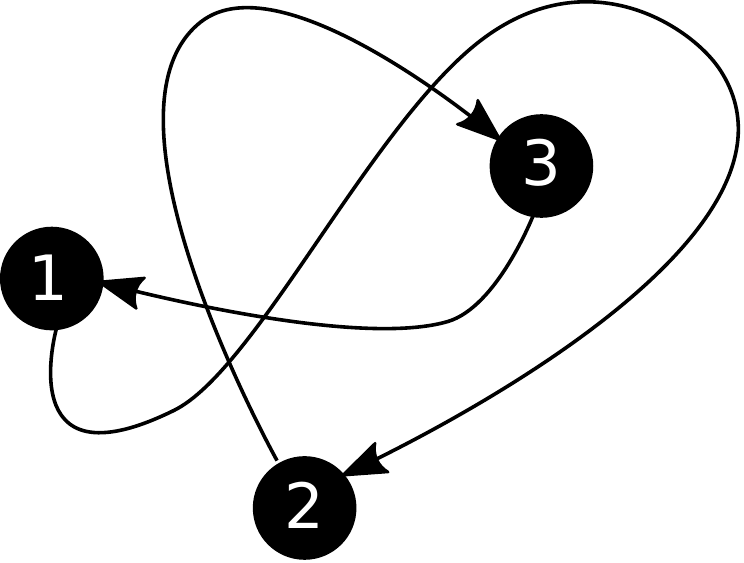}
\end{subfigure}
\begin{subfigure}[]{}
  \includegraphics[width=0.25\textwidth]{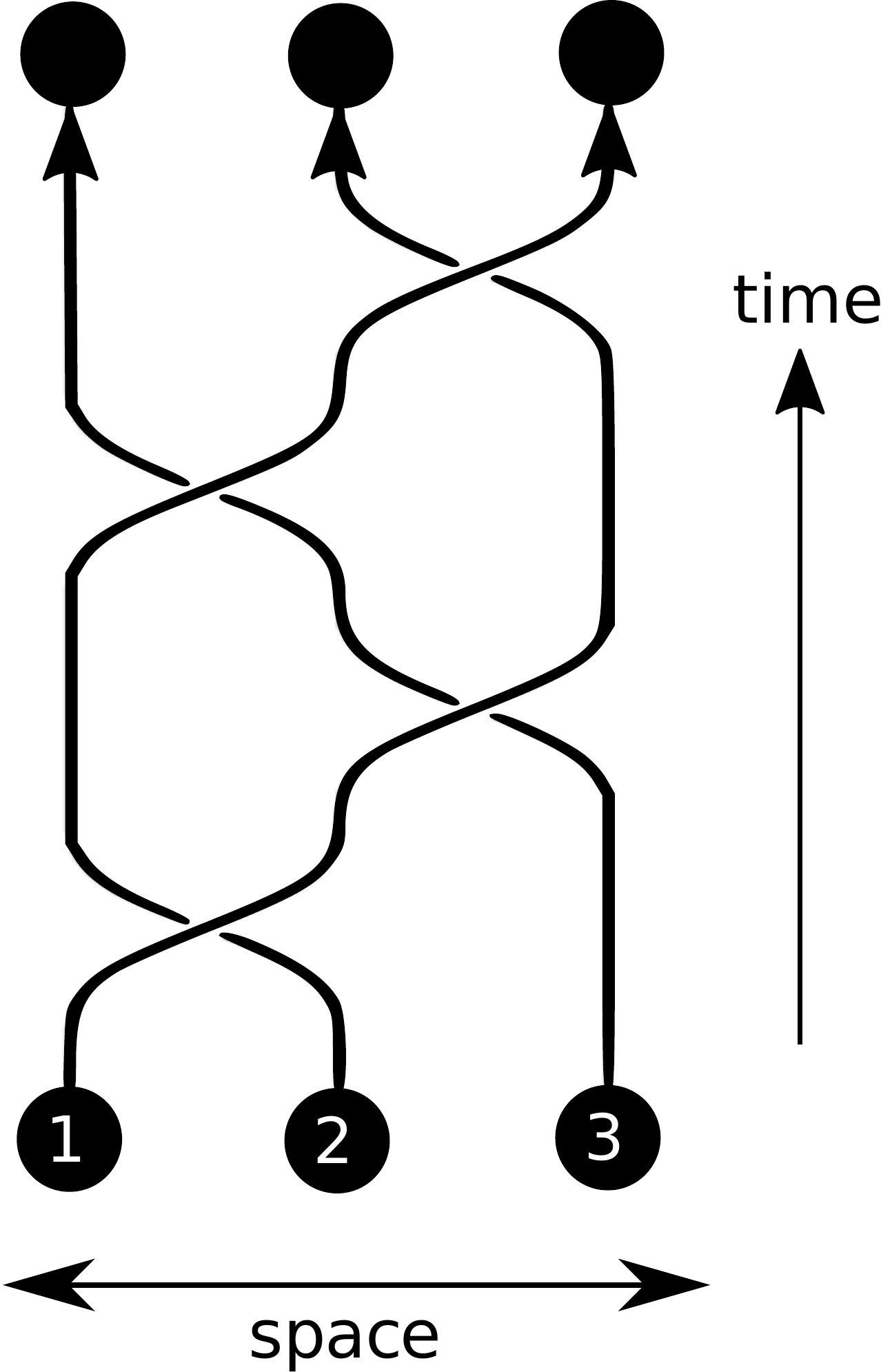}
\end{subfigure}
  \caption{Real-space trajectories of three particles (a), and the braiding of their worldlines (b). If the particles are non-Abelian anyons, then the final many-body quantum state is different than the initial one and information can be encoded as states that are related by different, topologically inequivalent braids. Deformation of the trajectories by noise or inaccurate control do not alter the stored information, as long as the particles do not come into contact. This robustness provides resilience to errors.}\label{braiding}
\end{figure}

Certain quantum states could also be braided along space and time dimensions, visualized in Figure \ref{braiding}. Superpositions of the occupation number of those states enable the storage of quantum information. Many interactions with the environment occur over small length scales, and are exponentially suppressed as the braiding is performed over larger distances. Only specific kinds of processes may still destroy the information. One such example is quasiparticle poisoning, where the environment introduces additional particles into the system. These particles might braid and fuse with the existing particles in the system, but the rate of this process is still estimated to allow for decoherence times on the order of seconds. \citep{PhysRevLett.126.057702}

The quantum states that possess the necessary properties describe particles called non-Abelian anyons. \citep{Stern2010} Elementary particles of this type have not been discovered in nature, but it could be possible to create quasiparticle excitations within low-dimensional solid-state systems, which fit into this specific class of particle. \citep{Wilczek2009} Recent experimental research shows evidence for the existence of anyons \citep{Bartolomei2020,Nakamura2020}, but so far there has been no conclusive observation of any non-Abelian anyons. Consequently, no topological computers or qubits have been built to date.

Current proposals involve fabricating semiconductor-superconductor hybrid nanostructures to engineer topological superconductors, which are theoretically predicted to host non-Abelian quasiparticles at their boundaries. \citep{Lutchyn2018,Lutchyn2010,Oreg2010} This approach has the potential to offer a path towards a scalable, fault-tolerant topological quantum computer. Similarly to superconducting qubits, extremely low temperatures are required for operation. \citep{Karzig2017,Kitaev2003}

\section{Quantum technology}
There are many different types of quantum technologies, with each type having its specific characteristics with different advantages and disadvantages. This section will focus on three of those technologies: superconducting circuits, trapped ions and photonic qubits.

\subsection{Superconducting circuits}
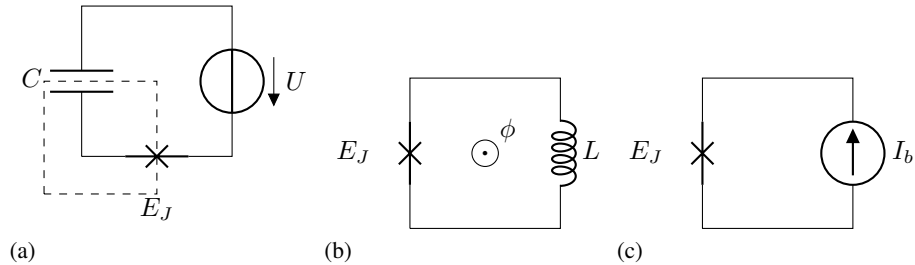
\begin{figure}[htbp!]
\begin{centering}
\begin{subfigure}[]{}
    \begin{circuitikz}
      \draw (0,0)
      to[C=$C$] (0,2) 
      to[short] (2,2)
      to[V,v=$U$] (2,0) 
      to[barrier,l=$E_J$] (0,0); 
      \draw[dashed](-0.5,-0.5)rectangle(1,1);
    \end{circuitikz}
\end{subfigure}
\begin{subfigure}[]{}
    \begin{circuitikz}
      \draw (0,0)
      to[barrier,l=$E_J$] (0,2) 
      to[short] (2,2)
      to[L,l=$L$] (2,0) %
      to[short] (0,0);
      \draw[black] (1,1) circle[radius=5pt] node[]{\textbf{.}};
      \draw[black] (1.3,1.3) node[]{$\phi$};
    \end{circuitikz}
\end{subfigure}
\begin{subfigure}[]{}
    \begin{circuitikz}
      \draw (0,0)
      to[barrier,l=$E_J$] (0,2)
      to[short] (2,2)
      to[american,I,invert,l=$I_b$] (2,0)
      to[short] (0,0);
    \end{circuitikz}
\end{subfigure}
    \caption{(a) Charge, (b) flux and (c) phase qubit}\label{figSuperconducting}
\end{centering}
\end{figure}

Superconducting qubits use energy levels of superconducting circuits as computational basis states. Based on the degree of freedom, three main categories have been proposed for this type of qubit \citep{kjaergaard2020superconducting}:
a) charge qubits, b) phase qubits and c) flux qubits.
\\
Two factors can be influential in determining these three categories: Josephson energy (JE), which describes the energy stored in a Josephson junction, and charging energy (CE), which describes the energy described in a capacitor.
The ratio JE/CE can distinguish the classes. In Figure \ref{figSuperconducting}, a) represents the charge qubit, which is also called the cooper pair box. In this case, the CE is larger than the JE (JE/CE \(<<\) 1). The basis states in the charge qubits are the charge states, which means the presence or absence of cooper pairs on the island. The state of the qubit can be changed by changing the number of cooper pairs that cross the junction.
\\
In Figure \ref{figSuperconducting}, b) depicts the flux qubit circuit diagram. In this case, the JE is larger than CE (1 \(<\) JE/CE \(<\) 100). The state of the qubit can be changed by changing the bias flux \(\Phi\).
\\
In Figure \ref{figSuperconducting}, c) depicts the phase qubit, which is a current-biased circuit where JE is much larger than CE. Accordingly, the JE/CE ratio is considerable larger than the ration for the charge qubit.\\
Additionally, different types of superconducting qubits can be derived by combining features of the three mentioned main types. For example, C-shunt flux \citep{yan2016flux}, Transmon \citep{koch2007charge}, and Hybrid qubits \citep{steffen2010high}.

\begin{table}[htbp!]
\caption{Advantage and disadvantage of superconducting qubits}
\label{table2}
\resizebox{\linewidth}{!}{%
\begin{tabular}{c|c}
Advantage                        & Disadvantage \\
\hline
Faster gate speed                & Poor connectivity to other qubits \\ 
The most mature of all the types & Requires near absolute zero temperature to operate \\ 
Can be built using existing semiconductor approaches & Susceptible to noises \\ 
Easy to reproduce qubits         &  Keeps qubit states for short period \\
\end{tabular}}
\end{table}

\subsection{Trapped ions}
The first quantum gate for the trapped-ion systems was proposed by Ignacio Cirac and Peter Zoller. \citep{cirac1995quantum} This was the first step to turn quantum computing from theory into an experimental issue. Today it is one of the leading candidates in quantum computing \cite{bruzewicz2019trapped}, with Honeywell's H1 system achieving the highest recorded quantum volume, which is a measure of performance capability for quantum computers \cite{cross2019validating}, in 2021. \cite{honey2021} This qubit technology uses ions, such as $\text{Yb}^{3+}$ \cite{kindem2020control}, suspended in vacuum, where certain internal electronic states of the ions are used as the computational basis states. Based on which states are chosen for this, four types of qubits are distinguished: hyperfine qubits \citep{blinov2004quantum}, zeeman qubits \citep{brown2018comparing}, fine structure qubits \citep{bruzewicz2019trapped}, and optical qubits \citep{bruzewicz2019trapped}. The most important difference of these types is the amount of energy by which the internal states differ, enabling (and requiring) different photon energies for control and readout, which in turn provides unique benefits and drawback to each sub-approach. The general advantages and disadvantages of trapped ion qubits are shown in Table \ref{table3}.

\begin{table}[htbp!]
\caption{Advantage and disadvantage of trapped ion qubits}
\centering
\label{table3}
\begin{tabular}{c|c}
Advantage & Disadvantage  \\
\hline
Stability (long coherence time) & Slow gate speed  \\ 
Optimal connectivity (all to all) & Many lasers are needed  \\ 
Entanglement of qubits is easy & Vacuum needed for operation \\
Near absolute zero temperature is not needed & \\
\end{tabular}
\end{table}

\subsection{Photons}
Quantum theory describes light as consisting of individual particles called photons, which can be used to carry quantum information. The aim of photonic quantum computing (PQC) is to embed the processing of quantum information on photons into electronics so that it can be used for quantum computation. To perform PQC, photons need to be generated and quantum information needs to be encoded onto them. Subsequently, the photons must be manipulated to perform the computation and then detected to retrieve the result. Each of these steps have unique requirements and challenges associated with them. \citep{slussarenko2019photonic, flamini2018photonic} The general advantages and disadvantages of PQC are diplayed in Table \ref{photontable}.

\begin{table}[htbp!]
\caption{Advantages and disadvantages of photonic quantum computing}
\label{photontable}
\resizebox{\linewidth}{!}{%
\begin{tabular}{c|c}
Advantage  & Disadvantage  \\
\hline
Extremely long coherence time    &  Multi-qubit gates hard to implement\\ 
Possible room temperature operation & Deterministic photon generation and detection difficult \\ 
Easy to generate entanglement & Qubits are destroyed by measurement \\
Easy to perform reliable single-qubit gates\\
Scalable architectures possible & \\
Supports quantum communication and distributed QC
\end{tabular}}
\end{table}

\subsubsection{Photon generation}
Photons can be generated easily and in high numbers. For example, a common light bulb emits $\sim10^{20}$ photons each second. The difficult task is to build a mechanism for deterministic single photon generation, which would be capable of emitting precisely one photon on demand, such that it is directed into an optical fiber or other specific element of the photonic circuit. Moreover, when using many photons within a computation, they need to be indistinguishable from each other in order for two-photon interference effects to give predictable and usable results.

Improving the reliability of deterministic single photon sources is an active area of research and many advancements have been made in recent years using trapped ions \citep{higginbottom2016pure}, colour centers \citep{benedikter2017cavity} and quantum dots \citep{lodahl2017quantum, uppu2020chip, dusanowski2019near}. A near-deterministic source could also be constructed by combining logical control operations on probabilistic processes. For example, spontaneous parametric downconversion (SPDC) probabilistically converts one higher energy photon into two lower energy photons. Detecting one of the photons heralds the other, which can then be routed into the circuit for computation. \citep{caspani2017integrated}

\subsubsection{Encoding information onto photons}
Information can be mapped onto photon states in numerous ways. A qubit can be constructed from photons by choosing two orthogonal photon states that represent the logical states $\ket{0}$ and $\ket{1}$ of the qubit. This is sometimes referred to as dual-rail encoding. For example, a photon with horizontal polarization could represent $\ket{0}$ and a photon with vertical polarization could represent $\ket{1}$. A single photon has the ability to be in an arbitrary superposition of these two states and could as such be used as a qubit. A different example would be to send photons through two possible paths and a photon going through one path would represent $\ket{0}$ while a photon going through the other path would represent $\ket{1}$. Again, photons can travel through a superposition of many paths, experimentally easily achievable with a beam splitter. Many other encoding schemes exist as well and multiple encodings can be used simultaneously, meaning a single photon could encode multiple qubits in different ways at the same time. \citep{kagalwala2017single} Moreover, some degrees of freedom are not just restricted to two possible states, but rather a plethora or even continuum of states, which could be used to encode \textit{qudits}, the higher-dimensional version of qubits. As an example, a single optical mode can be in a superpositon of Fock states $\ket{n}$, each of which represent a state with a certain number of photons $n$. There is, in principle, no limit to the amount of photons in a Fock state, allowing for an arbitrary amount of information to be encoded into one such optical mode, provided we have the capability to accurately prepare such states and distinguish between them.

\subsubsection{Computation on photons}
Perhaps the most difficult aspect of PQC is performing the actual computation itself. More specifically, the ability to perform multi-qubit operations is a necessary requirement for a universal quantum computer and for most quantum algorithms. The issue is that in normal circumstances, photons do not exhibit non-linear interactions with each other, which is good for avoiding unwanted cross-talk between qubits but detrimental to engineering multi-qubit gates. Early attempts to perform multi-qubit gates on photons involved using non-linear optical elements such as a Kerr media \citep{chuang1995simple}, but the viability of this approach is severely limited because this effect is extremely weak in all known materials. \citep{kok2002single}

Consequently, most of the effort in the field focusses on linear optical quantum computing (LOQC), which uses only linear optical elements and measurements. \citep{kok2007linear} An important advancement came with the discovery of the KLM scheme, which probabilistically performs a two-qubit gate using just these components. \citep{knill2001scheme} This scheme was quickly improved upon by others to perform almost deterministic two-qubit gates. \citep{kok2007linear} This shows that universal photonic quantum computing is in principle possible, even in the framework of LOQC.

A related, but conceptually different approach is to perform measurement based quantum computing on optical cluster states \citep{nielsen2004optical}, where the goal is to generate a pre-determined highly entangled network of qubits as the input and perform quantum computation by only using single-qubit gates conditioned on projective measurements. Recent experiments have demonstrated that it is possible to create optical cluster states with thousands of qubits \citep{larsen2019deterministic,asavanant2019generation} and perform deterministic multi-qubit operations on them. \citep{larsen2021deterministic}

Finally, a non-universal quantum computational scheme called boson sampling has been pursued actively in recent years. The basic task of boson sampling is to send a number of input modes through a linear interferometer and sample the distributions of the photons at the output. \citep{aaronson2011computational} One variant called gaussian boson sampling uses gaussian states of light as input rather than single photons, thereby circumventing the difficulties of developing reliable single-photon sources. \citep{hamilton2017gaussian} The task of simulating the sampled photon distribution with a classical computer has no known efficient algorithm, which is why researchers have been able to claim a demonstration of quantum supremacy based on this problem. \citep{zhong2020quantum} Numerous near-term applications for this computational scheme have been found and compact, scalable and programmable devices are being developed for broader commercial use. \citep{bromley2020applications,arrazola2021quantum}

\subsubsection{Photon detection}
Due to the bosonic nature of photons, many photons may occupy the same quantum state at once. Broadly, photon detectors can be classified into two groups: photon number resolving (PNR) detectors, which can discriminate between different photon occupation numbers, and threshold detectors, which can only signal whether or not at least one photon occupies the state. PNR capabilities are obviously preferred, but may not be necessary for some algorithms and experiments. Current state-of-the-art for experiments are silicon avalance photodiodes (APDs), operating in Geiger mode, which are threshold detectors with limited wavelength coverage. \citep{eisaman2011invited} Improving the capabilities of photon detection is an active area of research, with perhaps the most notable direction being superconducting nanowire single-photon detectors (SNSPDs). \citep{you2020superconducting} These detectors require cryogenic cooling to a few kelvin to enable superconductivity, which is still much more readily achievable than the millikelvin temperatures involved in superconducting qubits. Current state-of-the-art with PNR technology resides with transition-edge sensors (TESs), which are capable of resolving photon numbers but suffer from slow operation \citep{burenkov2017full,calkins2011faster}, roughly two orders of magnitude slower than what is considered practical for PQC. \citep{slussarenko2019photonic}

\subsection{Other quantum technologies}
There exist a range of other quantum technologies which we decided not to cover in this publication. These technologies are based on silicon quantum dots \cite{chatterjee2021semiconductor,ferraro2020all}, diamond vacancies or neutral atoms and the interested reader is referred to the referenced literature \citep{pezzagna2021quantum,henriet2020quantum,saffman2019quantum}.

\section{Bibliography Analysis}
\begin{figure}[htbp!]
\centering
  \includegraphics[scale=0.4]{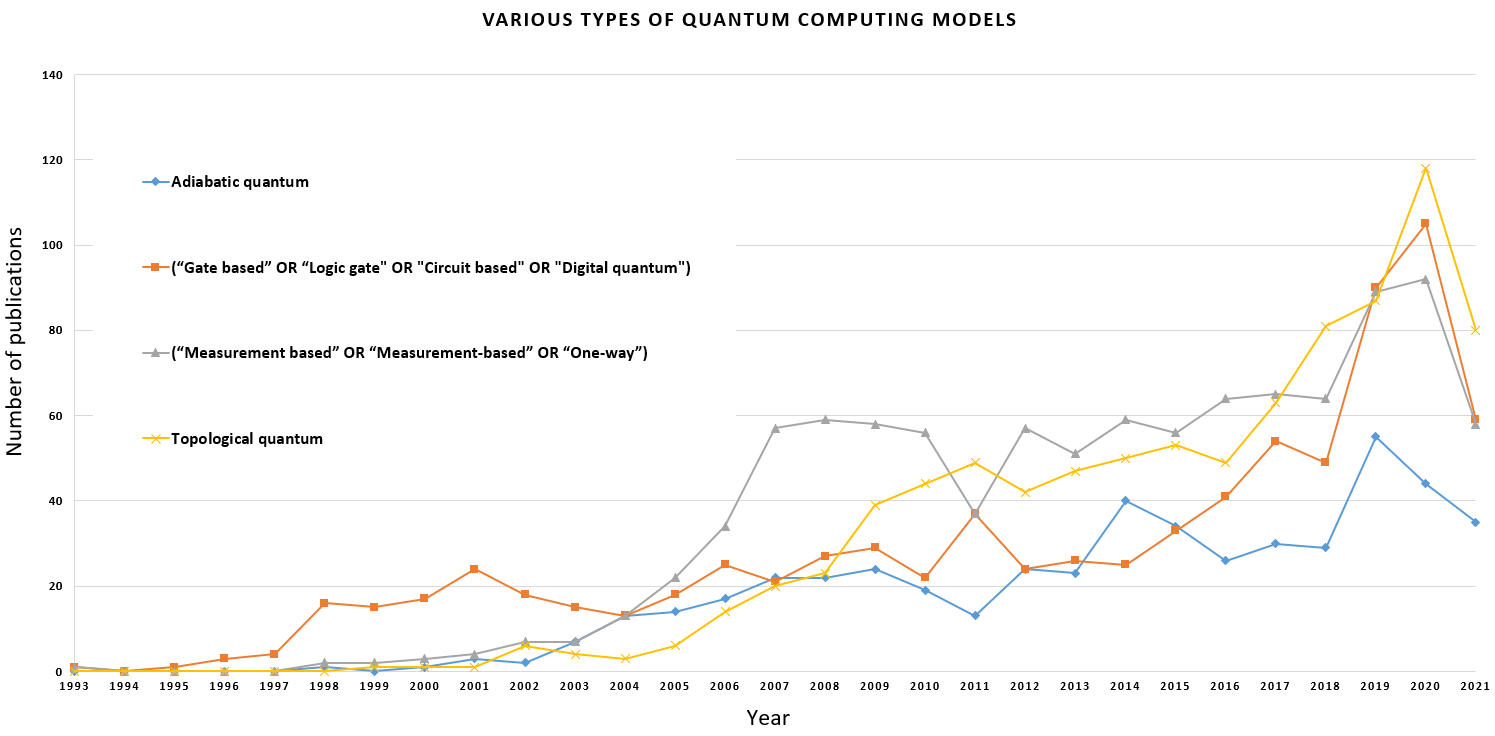}\\
  \caption{Search in topics from 1993 to 2021. Topics contain “quantum computing OR quantum computation” AND the keywords that are in the legend. These keywords are related to the various types of quantum models. }\label{figModels}
\end{figure}

In the past 30 years, quantum computing has experienced continuous expansion and diversification. This section describes trends and perspectives for quantum models, qubit technology, and the use of quantum computing in chemical and biochemical engineering. Publications are a suitable source of data that can be used as a factor to analyze the growth rates of scientific fields and for this bibliography analysis data is collected from Web of Science (WOS). 
Figure \ref{figModels} shows a quantitative analysis of published articles from 1993 to 2020. In this section, “search in topics” means search through the input fields for title, abstract, author keywords, and keywords. The figure is based on a list of composite keywords:
\begin{itemize}
\item (“quantum computing” OR “quantum computation”) AND “adiabatic quantum”
\item	(“quantum computing” OR “quantum computation”) AND (“gate based” OR “logic gate" OR "circuit based" OR "digital quantum")
\item	(“quantum computing” OR “quantum computation”) AND (“measurement based” OR “measurement-based” OR “one-way”)
\item	(“quantum computing” OR “quantum computation”) AND “topological quantum”
\end{itemize}
In total, 2,346 articles were found, from which 680 articles belong to the topological quantum model, 857 articles to the measurement-based models, 418 articles to the adiabatic quantum models, and 391 articles to the gate-based models.
The first published paper on quantum models is related to the gate-based model. \citep{mizutani1993novel} 
From 1998 to 2001, 11 articles have been found and classified as relating to measurement-based quantum computing models. \citep{castagnoli2001theory,castagnoli1993notions,fortnow1999complexity,castagnoli1998quantum,fortnow1999complexity,csurgay2000toward,yamasaki2000one,ambainis2000private,handel2001quantum,servedio2001separating} However, based on citations, paper \citep{raussendorf2001one}, with 3,918 citations, is the most significant paper on this model. This paper discusses one-qubit measurements of cluster states.
The first publication on the topological quantum model was in 1998. \citep{freedman1998p}

\begin{figure}[htbp!]
\centering
  \includegraphics[scale=0.4]{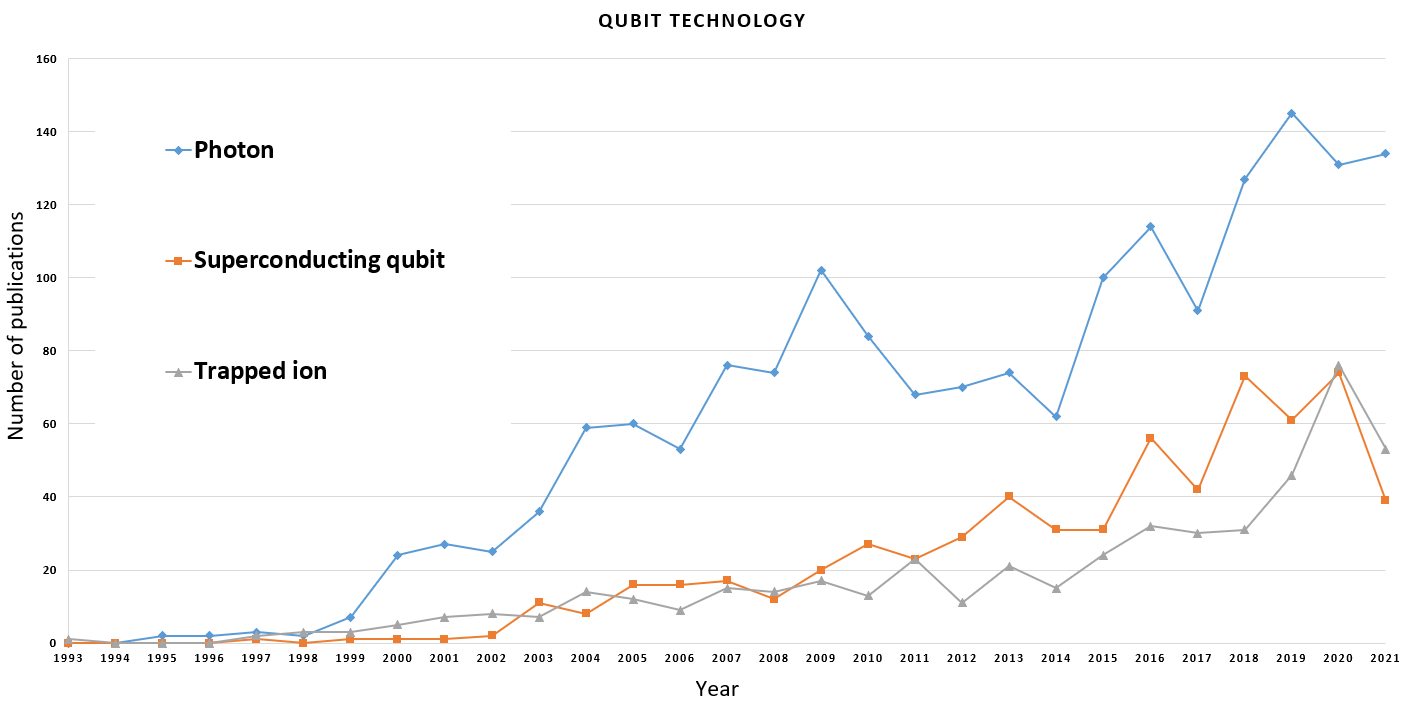}\\
  \caption{Search in topics from 1997 to 2020. Topics contain “quantum computing/computation” AND the keywords that are in the legend. These keywords are related to the various types of qubit technologies.}\label{figTech}
\end{figure}

During the 90s, the gate-based models were at the top of the chart (from 1993 to 2004). Then the number of published papers on this model gradually decreased. From 2004 to 2011, measurement-based models were consistently at the top of the chart.
Currently, the number of publications on topological quantum models and measurement-based models are ahead. In the last decade, 1,639 articles were published, of which 588 articles relate to the measurement-based model, 562 articles to the topological models, 282 to the adiabatic model and 197 to gate-based models. 

Figure \ref{figTech} illustrates a quantitative analysis for identifying trends throughout the years for two main types of qubit technology, trapped ion and superconducting qubit technology. The vertical axis shows the number of publications and the horizontal axis shows years, from 1997 to 2020. The figure is based on a list of composite keywords:
\begin{itemize}
\item (“Quantum Computing” OR “Quantum Computation”) AND “Superconducting”
\item	(“Quantum Computing” OR “Quantum Computation”) AND (“Trapped ion")
\end{itemize}
The figure contains two lines such that each line shows one of the models described in Section 2. In total, 872 publications were found, of which 356 articles belong to trapped ion technology and 516 articles to superconducting qubits.

\begin{figure}[htbp!]
\centering
  \includegraphics[scale=0.55]{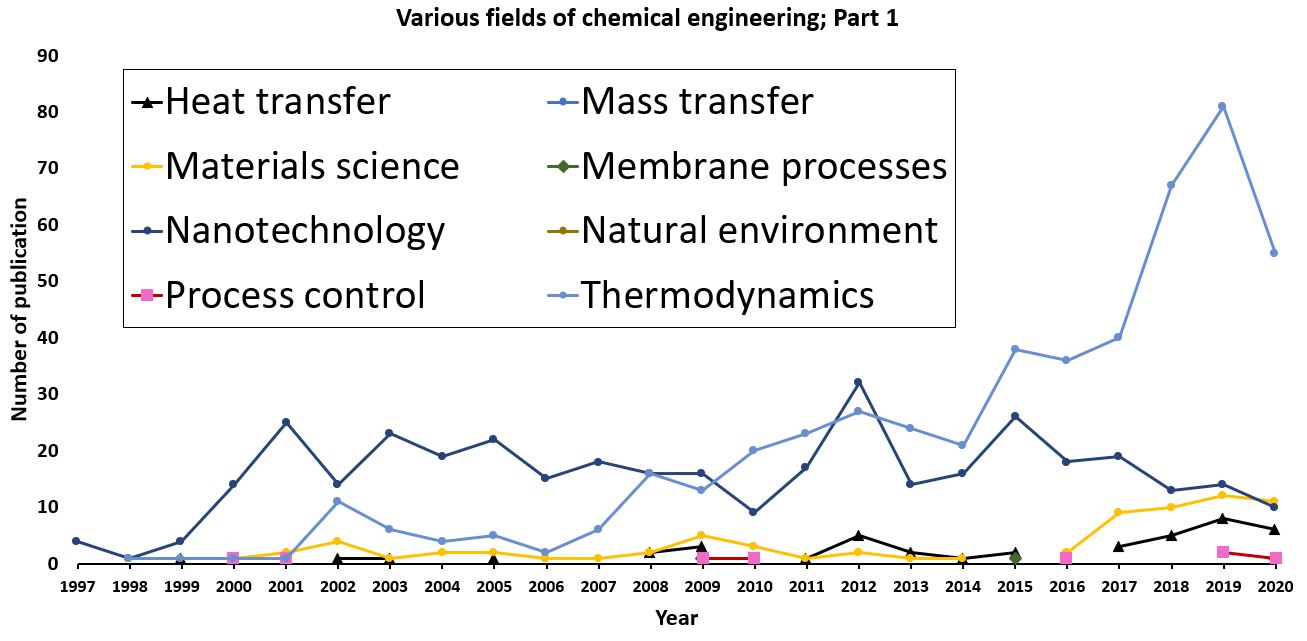}
  \caption{Search in topics from 1992 to 2020. Topics contain “quantum computing" OR "quantum computation” AND the keywords that are in the legend. These keywords are related to the various fields of chemical engineering part 1.}\label{figChem2}
\end{figure}

\begin{figure}[htbp!]
\centering
  \includegraphics[scale=0.55]{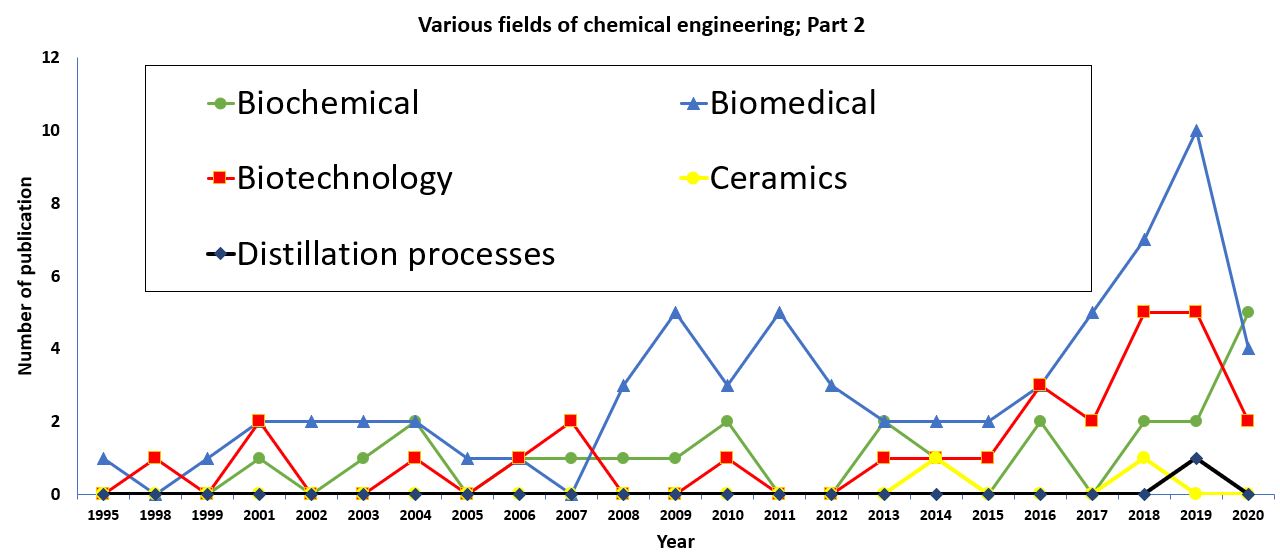}
  \caption{Search in topics from 1995 to 2020. Topics contain “quantum computing/computation” AND the keywords that are in the legend. These keywords are related to the various fields of chemical engineering part 2.}\label{figChem1}
\end{figure}

\begin{figure}[htbp!]
\centering
  \includegraphics[scale=0.55]{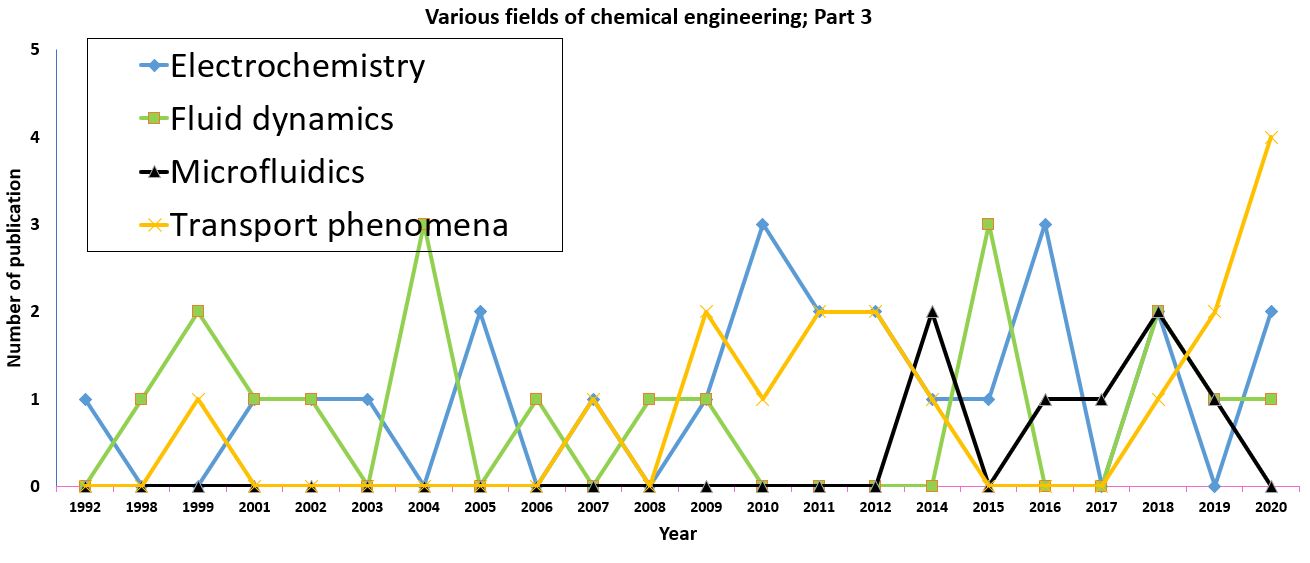}
  \caption{Search in topics from 1992 to 2020. Topics contain “quantum computing/computation” AND the keywords that are in the legend. These keywords are related to the various fields of chemical engineering part 3.}\label{figChem03}
\end{figure}

\begin{figure}[htbp!]
\centering
  \includegraphics[scale=0.7]{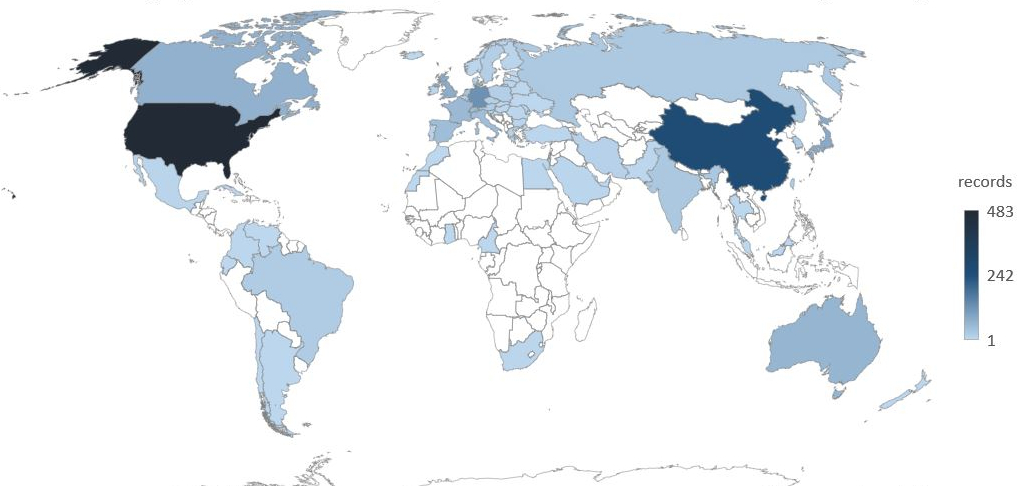}
  \caption{Search distribution world-wide from 2012 to 2020. Search contains (“quantum computing OR computation”) AND ("chemical" OR "chemistry" OR "biochemical")}\label{figMap}
\end{figure}

Figure \ref{figChem1} shows the result of the search in topics from 1997 to 2020. This figure illustrates the number of papers that are related to the sub-fields of chemical engineering (heat transfer, mass transfer, material science, membrane process, nanotechnology, natural environment, process control, thermodynamics) and QC.
In total, 923 papers were found, of which 444 belong to the field of thermodynamics, 369 to nanotechnology, 62 to material science, 36 to heat transfer, 7 to process control, 2 to natural environment, 2 to membrane process, and 1 to mass transfer.
In 1997, four papers were published, all with respect to nanotechnology. \citep{young1997unraveling,porod1997quantum,draganescu1997solid,bandyopadhyay1997switching}
The first paper in the thermodynamics field was published in 1998. \citep{ulyanov1998physical} The authors tried to describe the principles of quantum computing and quantum search algorithms by considering new informational technologies and modern physics. "Modern physics" (at that time) is described in two main sections. The first section is quantum information theory and the second is non-equilibrium thermodynamics and quantum mechanics.
Figure \ref{figMap} represents the country related distribution of QC related publication data in chemical and biochemical engineering. The data is collected based on the number of publications from 2012 to 2020. The keywords used are: 
\begin{itemize}
\item (“Quantum Computing” OR “Quantum Computation”) AND ("chemistry" OR "chemical" OR "biochemical").
\end{itemize}
Overall, 1157 articles were found. The top 10 countries based on the number of publications are the USA (483 articles), China (257), Germany (125), England (101), Japan (81), Canada (67), Australia (61), France (53), Spain (46), and Switzerland (46).

The assessment of the bibliography analysis represents that the application of QC in thermodynamics, biomedical, and biochemical are the most explored. These fields have moved beyond the testing phase, and practical applications are becoming more common. There is also a clear trend in qubit technologies (superconducting and trapped-ion), measurement-based model, and topological quantum modelling.

\section{Quantum computing in chemical and biochemical engineering}
Chemical and biochemical engineering \(C\&BC\) intersect the topics of chemistry, physics, biology and mathematics to serve the societal need of delivering industrial scale solutions to manufacture products and to generate energy carriers. QC will serve as a computing architecture to improve on existing methods to model quantum mechanical systems and solve them more efficiently and accurately in combination with classical algorithms. QC will allow specific optimization problems to be solved with exponential speed up in comparison to existing classical algorithms when the problem scales in size and lies within the NP complexity class and the bounded-error quantum polynomial time (BQP) class.
\\
Within the multi-scale layer concept of \(C\&BC\) and process systems engineering (PSE), quantum mechanical calculations can be applied for predicting the properties of chemical compounds and the kinetics of reaction networks. The prediction values are needed to design products and processes.
QC can potentially also contribute to the calculation of transport phenomena and fluid dynamics. \citep{Gaitan2020, Budinski2021, Ray2019}
This would lead to generating results faster since fluid dynamic problems can take weeks to solve on high performance computing platforms.
\\
Finally, it would be remiss not to consider the logistics for \(C\&BC\) products in the post-production stage. Here, channeling the right product to the right destination, in the right quantities at the right time, is necessary to prevent disruptions in a distribution chain. These kind of supply chain optimization problems can also be solved efficiently with QC.
\\
The current authors have taken the multi-layer view of \(C\&BC\) proposed by \citep{Gani2020} and reduced it to a simplified model (Figure 13) for the purpose of the current discussion.
\begin{figure}[htbp!]
\centering
  \includegraphics[width=0.45\textwidth]{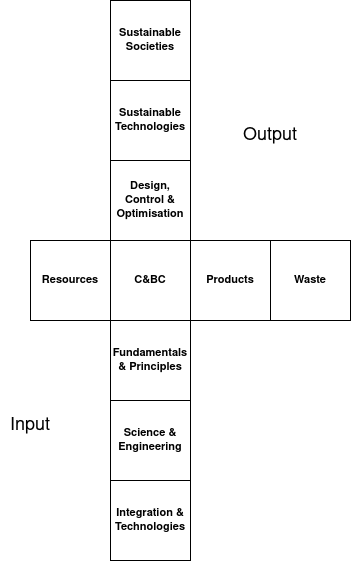}\\
  \caption{Simplification of C\&BC multi-layer view by Gani et al.}
\end{figure}

The final derived model (Figure 14) illustrates the relationship between five distinct classes which can be summarized under the headings:
\begin{itemize}
\item Resources
\item Knowledge \& work force
\item Quantum technologies
\item Products \& services
\item Market
\end{itemize}
Further, each class can be utilized in the creation of a landscape for QC in \(C\&BC\), where QC itself is an instance situated in the quantum technologies class.
Resources are needed in form of materials, compute power (sustained by electricity and cooling) and documented knowledge repositories. These resources are needed to develop quantum technologies, run the developed hardware for market usage and to educate students with the needed knowledge to perform research or work in the quantum technology domain.
\\
Figure 15 depicts the impact which quantum technologies and QC will have on sustainable technologies and how this will affect our societies in the world.
Near tearm quantum computing applications such as quantum machine learning, quantum chemistry and quantum Monte Carlo simulations will allow to accelerate the development of digital solutions and new materials such as catalysts or pharmaceutical drug compounds. We expect that quantum computing will contribute to improved energy efficiencies with respect to less electricity consumption of machine learning and optimisation algorithms. This leads consequently to a reduced carbon emission footprint. Further, the QC enhanced design of sustainable materials, processes and products will also have a beneficial impact on carbon emissions and other societal needs. The sections that follow will explore the opportunities for QC in the field of C\&BC and its' applications within the following:
\begin{itemize}
\item Pharmaceutical applications
\item Biochemical applications
\item Solid material applications
\item Process \& product design, optimization, monitoring \& control
\item Supply chain optimization
\end{itemize}

\begin{figure}[htbp!]
\centering
  \includegraphics[width=1.0\textwidth]{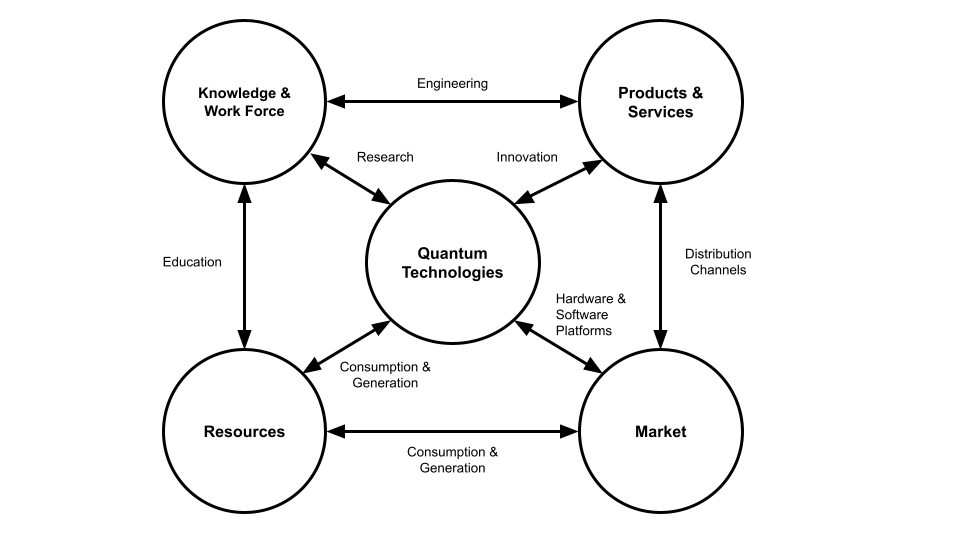}\\
  \caption{Relationship diagram for quantum technologies}
\end{figure}

\begin{figure}[htbp!]
\centering
  \includegraphics[width=1.0\textwidth]{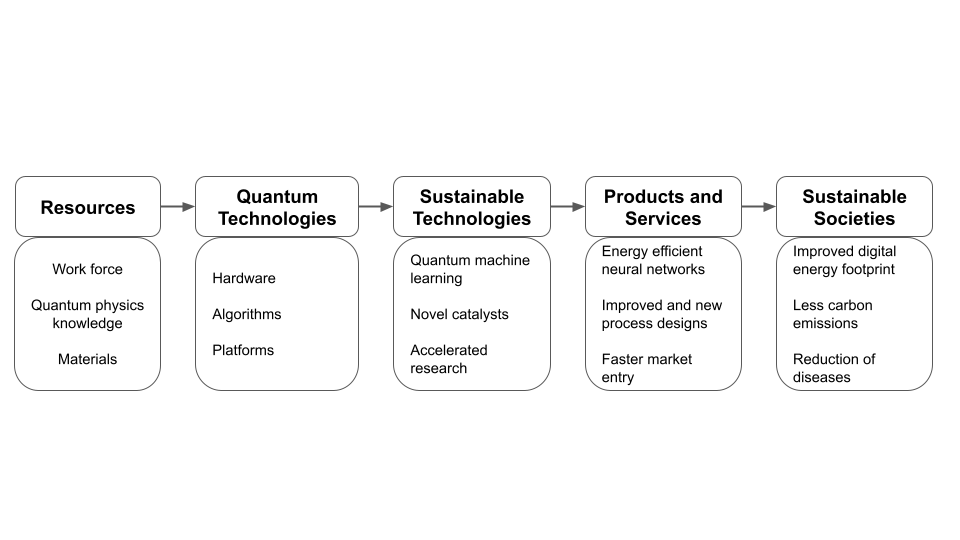}\\
  \caption{Impact of quantum technologies on sustainability}
\end{figure}

\subsection{Pharmaceutical applications}
The development of a market ready pharma product can take up to 12 years or longer \cite{Gaudelet2021}, and is characterized by tedious lab (in-vitro and in-vivo) experiments thus leading to high time investment and costs. Especially in the last decade in-silico experiments have been more and more integrated in the drug discovery pipeline due to the availability of high-quality structured data, next-generation sequencing technology, advances in quantitative biology and increased computer power. \citep{Sormanni2018, Ringel2020}
\\
Especially the drug discovery and pre-clinical research phases can benefit from quantum computing enhanced quantum chemistry calculations. \citep{Lam2020, Zinner2021, Cova2022} Amongst the recent developments in this area are big-pharma collaborations with quantum computing software and algorithm developers.
\\
For example Roche and Cambridge Quantum Computing (CQC), a UK-based developer of quantum computing software, are collaborating on designing quantum algorithms for early-stage drug discovery and development. \cite{CQC_Roche} A partnership between Seeqc and Merck has been formed to develop hybrid quantum-classical algorithms for commercial scale development. \citep{Seeqc_Merck} Zapata Computing is a Boston based startup which has successfully attracted an investment from BASF to assist in the development and discovery of new chemicals, pharmaceuticals and materials using QC. \citep{BASF_Zapata} These are only a selection of many examples to show that QC has attracted great interest in industry and several other announcements have been made between specialized quantum computing SMEs and big pharma companies.
\\
According to an article in C\&EN News, big pharma executives are discussing QC for drug discovery on a landscape where data-dependent discovery is dominating. \cite{Mullin2020}
The start-up MentenAI has focused for example on accelerating protein design with machine learning and quantum computing on the D-Waves architecture. \citep{Maguire2021}
Other use-cases in the drug discovery area can be found with respect to the modelling of chemical reactions and understanding how pharmaceutical drugs interact with proteins and enzymes in the human body. \citep{cao2018potential}
\\
Given all these developments, there still exist a great opportunity for exploring the application of QC where chemical engineers possess extensive know-how, namely pharmaceutical process development, control, optimization and monitoring. These areas involve a combination of skills within process systems engineering. Thus, QC will have an important contribution in multi-scale and enterprise-wide developments for integrated product-process development as further elaborated in section 5.4.
\\
Quantum machine learning can contribute to reducing the energy footprint of artificial intelligence (AI) and machine learning (ML) models based on neural networks. \cite{Andersson2021} Here, the training of these model can take several weeks and consume a high amount of electricity and thus contribute to CO2 emissions. Studies show that the training of a natural language processing (NLP) model emits more CO2 than the average of two US American citizens produce per year \cite{Strubell2020}, and the breakthrough Alpha Fold 2 model had to be trained for three weeks to win the protein folding contest CASP14. \cite{Jumper2021}
Ising models and thus quantum annealing can be applied to speed up various algorithms and it has been mathematically proven that convolutional neural networks can benefit from being trained on quantum compute devices with large data sets. \cite{Lucas2014, Pesah2021}
\\
In molecular design, a quantum computer can be used to design molecules and solid materials by considering specific properties, which leads to reduced lab work. Prediction of the molecular properties is one of the significant parts in molecular design. 
As \(C\&BC\) deals with the atomic scale, interactions are applied by the quantum behavior of nuclei and electrons. It is very hard for classical computers to model the quantum mechanical systems accurately enough because the interactions between particles are increasing very quickly as the system becomes more complex. \citep{aaronson2009quantum}
There are many computational tools that are being used by researchers for chemical computations, such as density functional theory (DFT) \citep{geerlings2003conceptual}, which provides an approximation of molecular systems.
However, these tools are effective for small molecules and have some limitations when dealing with large molecules such as proteins. But quantum algorithms have already been implemented as proof of concepts such as protein folding to be applied when the number of available qubits is large enough for real use-case problems. \citep{Perdomo-Ortiz2012}
To conclude, plenty applications of QC in chemical engineering exist with respect to pharmaceutical applications.

\subsection{Biochemical applications}
One of the very popular examples being mentioned in general articles about quantum computing applications is unveiling the detailed bio-synthetic reduction mechanism of dinitrogen to two ammonia molecules by the Mo-dependent nitrogenase enzyme, a complex metalloprotein. The most active site of this metalloprotein is the iron-molybdenum cofactor (FeMo-co). \citep{Reiher2017} A Fe and the MoFe protein form the pocket in nitrogenase where FeMo-co is contained. \citep{Seefeldt2009} This enzymatic process takes place at ambient pressure and temperature and if decoded could substitute the energy intensive Haber-Bosch process.
Reiher et al. \citep{Reiher2017} propose to calculate the reaction mechanisms by implementing a multi-configurational wave function as a sub-model to be solved on a quantum computer to improve/enhance classical methods such as density functional theory (DFT).
Apart from enzyme design \cite{Cheng2020}, these kind of computational studies of reaction mechanisms and kinetics can be found for a wide area of high impact applications such as green catalysis of the Mannich reaction \cite{Stevens2017}, biochemical redox reactions \citep{Jinich2019} and calculating uncertainties in microkinetic models \cite{Becerra2021}.

\subsection{Solid material applications}
Quantum computing could benefit combined ML and DFT models to for example analyze new battery materials where oxidation potentials have to be analyzed via high throughput screening. \citep{Doan2020} As already stated before machine learning itself can profit from quantum computing \citep{Zlokapa2021}, and further improve quantum chemistry calculations with ML approaches such as presented with FermiNet \citep{Pfau2020} or PauliNet \citep{Hermann2020}.
With respect to quantum chemistry calculations being one of the first domains to be applied on NISQ devices, novel battery systems such as calcium battery electrolytes can be analyzed more accurately and efficiently. \citep{Araujo2021}
Screening for synthetic catalysts can also be achieved by applying a hybrid classical-quantum algorithm where the free energies of all species in a reaction scheme are calculated on a quantum computer. This iterative process of providing initial parameter values of the Hamiltonian to the quantum algorithm to calculate the free energies for reaction pathway analysis will allow to modify the catalyst structure in each step to identify the most thermodynamic favorable reaction route. \citep{vonBurg2021}
\\
The analysis of polymeric structures allows to design more efficient plastic manufacturing and recycling processes \citep{Walker2020, Sanchez-Rivera2021}, and the design of novel bio-degradable polymers via redox switchable catalysis for example. \cite{Upton2014, Deng2021} These topics can be tackled with quantum chemistry and quantum computing in the future with highly automated algorithms. \cite{Deglmann2015}

\subsection{Process and product design, fault-diagnosis and logistics optimization}
One important aspect worth highlighting is that knowledge transfer from different disciplines and application fields within quantum computing will become an important part of future research to identify novel solutions within manufacturing.
One example is the work by Castaldo et al. where the optimal control in laser-induced population transfer is studied on the molecule cyanidin. \cite{Castaldo2021} This kind of work is important e.g. for dye-sensitized solar cells where the efficiency and stability of photovoltaic materials is still undergoing a phase of improvement before market readiness. \cite{Castillo-Robles2021} Castaldo et al. developed a hybrid quantum-classical algorithm involving also machine learning to simulate and control a ultra-short laser pulse directed at a cyanidin molecule to study the time-evolution of its state. Different algorithms were benchmarked and involved Broyden–Fletcher–Goldfarb–Shanno (BFGS), a genetic algorithm (GA) and Nelder–Mead (NM) methods combined with a quantum routine. These hybrid algorithms were compared against the classical Rabitz algorithm.
\\
This quantum optimal control theory (QOCT) example shows that quantum computing can tackle use-cases and implement algorithms which intersect material/product design, control, monitoring, optimization and in the end can also contribute to needed knowledge of designing or modifying a manufacturing process.
To leverage the potential of quantum computing, chemical engineers have to be involved to apply their systems thinking approach with the expertise of quantum physicists while abstractions have to be made so that all engineering disciplines are able to make us of quantum computers.
\\
First-principle models are important to design processes or products to accommodate the next industrial shift to Industry 5.0 applications. Industry 5.0 envisions a connection between customer defined products and the possibility that manufacturing plants are able to produce the defined product specification of the customer. Here we see the potential that materials can be designed via hybrid classical-quantum algorithms and manufacturing processes will be set with the specific process parameters to produce the specified product. Quantum computing can enhance process and product design pipelines which combine various computational models ranging from the property prediction layer to process and products models such as unit operation models or product formulation models.
\\
One such example, is designing solvents for various applications. Solvents are used in abundance in the chemical and biochemical industry. The use of efficient model-based solvent selection techniques is an option worth considering for rapid identification of candidates with better economic, environment and human health properties. The underlying solvation models which make use of quantum chemical calculations can benefit from quantum computing when methods have been developed to map such calculations to a quantum computer.
For example, several research publications have developed frameworks combining machine-learning and computer-aided molecular design (CAMD) where machine learning algorithms have been combined with molecular descriptor packages while quantum chemical calculations were performed with DFT and solvation models. \citep{Liu2021, Doan2020} In these frameworks, solving the density functional theory calculations using quantum computing can improve the accuracy and computational efficiency of these frameworks in the future.
\\
Several research papers have been published on the topics of manufacturing operations management \citep{Rieffel2014}, fault diagnosis \citep{Ajagekar2020a,Ajagekar2021a,Ajagekar2021b}, scheduling \citep{Rieffel2014,Tran2016}, logistics \citep{Neukart2017} and optimization of energy systems \citep{Ajagekar2019,Ajagekar2020b}.
\\
In order to address the computational challenges, hybrid QC-based algorithms are proposed in the referenced literature and extensive computational experimental results are presented to demonstrate their applicability and efficiency.
\\
Navigation-type and scheduling-type problems have been tackled with the D-Waves quantum computer by Rieffel et al. \citep{Rieffel2014}
The authors took two approaches to tackle these kind of problems: (I) to map from general classical planning problem formulations to QUBO form; (II) to look at the problems specifically and perform a direct mapping to QUBO form.
The general mapping was performed with two variants: a conjunctive normal form (CNF) instance and a time-slice instance adapted from \citep{Smelyanskiy2012}.
Direct mapping of graph coloring and direct mapping of Hamiltonian path formulations to the QUBO formulation were performed for the individual problem types. The direct mapping schemes have shown better performance than the more general undirected graph mapping scheme.
The conclusion is that direct mapping schemes for the individual specific problem at hand (navigation-type, scheduling-type) will outperform general-purpose mapping methods which try to cover multiple problems. Thus, the most beneficial would be to develop a methodology or framework which applies direct mapping schemes for specific planning (navigation-type, scheduling-type) problems.
\\
The attempts on tackling dynamic optimization and process monitoring by combining deep learning approaches and QC, a deep learning based fault diagnosis method and its application on selected problems such as continuous stirred tank reactors (CSTRs) and the Tennessee Eastman process has been proposed. \citep{Ajagekar2020a,Ajagekar2021a,Ajagekar2021b}
This is a domain where hybrid machine-learning and quantum assisted algorithms can be developed for process monitoring, fault diagnosis and control. \citep{Ajagekar2020a} However, to date such approaches have been only applied to benchmark simulation models and their real use-case application in this sector requires further research.
\\
A tree-search based quantum-classical algorithm has been implemented to solve a scheduling problem by dissecting the problem into a master formulation and multiple sub-problems. \citep{Tran2016} The master formulation was implemented on the quantum computer while the global tree search algorithm and the sub-problems were run on a classical device. This kind of concept bears similarity to the formulation of a molecular conformation search problem where also this kind of hybrid split up between master and sub-routines between quantum and classical devices has been performed. \citep{Ajagekar2020b}
A flexible job shop scheduling problem has been developed with a quantum-inspired quantum annealer and digital annealer algorithms. \citep{Venturelli2016,Denkena2021} The quantum annealer algorithm was implemented as a time-indexed QUBO problem derived from a makespan-minimization problem formulation. The digital annealer implementation extends the before mentioned QUBO formulation with a penalty term to adjust for shorter makespans in the schedule.
\\
The computational power plays a significant role in scheduling and logistics applications, especially when applying machine learning and considering multiple variables. The number of variables can increase because of real-time applications and market demands. For example, the number of constraints on the system can be affected by the out-of-stock products or fleet breakdown. Decision making is at the core of such supply chain systems to reduce operational costs. \citep{exploringibm2019} Fujitsu and Toyota proposed optimizing supply chain and logistic network operations using Fujitsu’s quantum-inspired digital annealer computing solution. \citep{fujitsu2020} The results showed that it takes 30 minutes to determine an optimal route that can reduce the logistics cost by about 2 to 5 percent.
\\
Not too much work has been published yet specifically dealing with process control theory and quantum computing. But we expect in the upcoming years that more hybrid classical-quantum deep/machine learning algorithms will be applied to controlling unit operations or entire processes. Ajagekar et al. \citep{Ajagekar2020a} have developed a deep learning algorithm to control a CSTR and the Tennessee Eastman process. Restricted boltzmann machines (RBMs) were implemented to generate via quantum sampling the run data for the quantum annealer which delivers then the expectation values to set the model parameters.
The CSTR fault diagnosis problem shows improved performance with the QC algorithm with respect to fault detection rates.
In case of the Tennessee Eastman process the deep learning algorithm adapted from \citep{Zhang2017} showed superior performance over the principal component analysis (PCA) algorithm whereas the QC adapted deep learning algorithm showed better fault detection rates for some individual detected faults and some for rates equal to 0 for the classical deep learning algorithm.
\\
We also expect that quantum machine learning (QML) will have an impact on chemical and biochemical process control applications since many machine learning algorithms are being developed for process control.
Important for QML algorithms is the storage of the input data on the random access memory (RAM) of the quantum device termed QRAM. And research has already been performed to implement quantum-inspired regression to benchmark against quantum algorithms. \citep{Gilyen2020,Tang2021} It was also shown under which practical conditions quantum-inspired and quantum algorithms can achieve practical results. \citep{Arrazola2020} Quantum-inspired machine learning is suitable for input matrices with a low rank, low condition number and a very large dimension whereas quantum machine learning algorithms can be fed with sparse matrices of a high rank.
Further to highlight with respect to QML is that a mathematical proof has been delivered showing that convolutional neural networks (CNNs) can be solved at scale with large data sets as a quantum CNN (QCNN) in contrast to classical CNN implementations. \citep{Pesah2021}
\\
Ajagekar et al. \citep{Ajagekar2019,Ajagekar2020b} have solved unit commitment, facility-location, and heat exchanger networks problems using QC. In their work, they have concluded that the current capacity of quantum computers negatively affects the quality of solutions obtained for large-scale problems. However, this can be overcome in the next few years as the number of available qubits will increase and error-correction approaches will increase the performance of quantum computers.

\section{Conclusion}
This paper summarizes the fundamentals of QC, quantum computers, the application of QC in \(C\&BC\) engineering and future progress directions. The main goal of this paper is to provide an overview to chemical and biochemical researchers and engineers who are not yet familiar with quantum computation.
\\
Numerous articles were dedicated to the different applications of QC in \(C\&BC\) engineering. The result of the bibliography analysis showed a 10.29 percent growth in 2020. These statistics in chemical and biochemical topics equal 1,010 papers out of 34,698 with 43.51 average citations per paper, where the average citation per year equals 1569.64.
\\
The paper provides a detailed view into quantum computing hardware and conducted research in \(C\&BC\) engineering. The intention is that the reader picks up an interest to read up on one of the applications or quantum hardware devices and hopefully starts re-implementing their own algorithms and further develop them.

\bibliographystyle{unsrtnat}

\end{document}